\documentclass[12pt,letterpaper]{article}

\usepackage{amsmath,amssymb,calc}
\usepackage{graphicx}

\newcommand\be{\begin{equation}}
\newcommand\ee{\end{equation}}
\newcommand{\bea}{\begin{eqnarray}}
\newcommand{\eea}{\end{eqnarray}}

\newcommand{\la}{\langle}
\newcommand{\ra}{\rangle}

\def\id{\protect{{1 \kern-.28em {\rm l}}}}

\def\unit{\relax{\rm 1\kern-.26em I}}

\def\id{\protect{{1 \kern-.28em {\rm l}}}}

\begin{document}

\begin{titlepage}
\begin{center}
\hfill QMUL-PH-08-08 \\
\hfill ITP-UU-08/16 \\
\hfill SPIN-08/16 \\
\vskip 15mm

{\Large {\bf Finite Temperature Behaviour of O'KKLT Model\\[3mm] }}

\vskip 10mm

{\bf Lilia Anguelova$^{a}$ and Vincenzo Cal\`o$^{b}$}

\vskip 4mm
$^{a}${\em Institute for Theoretical Physics and Spinoza Institute}\\
{\em Utrecht University, 3508 TD Utrecht, The Netherlands}\\
{\tt L.Anguelova@phys.uu.nl}\\[2mm]
$^b${\em Center for Research in String Theory}\\
{\em Department of Physics, Queen Mary, University of London}\\
{\em Mile End Road, London, E1 4NS, UK.}\\
{\tt v.calo@qmul.ac.uk}\\

\vskip 6mm

\end{center}

\vskip .1in

\begin{center} {\bf Abstract }\end{center}

\begin{quotation}\noindent
We summarize our work on the O'Raifeartaigh uplifted KKLT model at finite temperature. We study the system for parameter values for which the zero temperature potential for the volume modulus has a dS minimum. The temperature-dependent part of the effective potential has a runaway behaviour for one exponent in the superpotential, whereas it can have local minima at finite field vevs for two exponents in the superpotential. However, it turns out that, despite the presence of those minima, the zero temperature dS vacuum is not destabilized by thermal corrections within the whole range of validity of our approximations.
\end{quotation}
\vfill

\end{titlepage}

\eject


\section{Introduction}

Moduli stabilization is a major problem on the road to relating string theory compactifications with phenomenology. It stems from the fact that the internal space in purely geometric compactifications can have various deformations, which manifest themselves in the four external  dimensions as scalar fields without a potential energy (called moduli). As the 4d effective action depends on those moduli, allowing arbitrary values for the latter leads to a lack of predictability of various 4d parameters (coupling constants) and a huge vacuum degeneracy. To resolve this problem, one has to turn on background fluxes \cite{DRS,GKP} and, in the case of type IIB, also take into account non-perturbative effects. The fluxes generate a superpotential for all geometric moduli in type IIA on CY(3) \cite{IIA} and for the complex structure moduli in type IIB \cite{GKP}. So for the latter case, one needs to include also non-perturbative effects \cite{KKLT} (or a combination of perturbative and non-perturbative corrections \cite{Quevedo}) in order to stabilize the K\"{a}hler moduli.

Type IIB flux compactifications are better understood, compared to the type IIA ones, because there is a class of solutions such that the backreaction of the fluxes on the geometry is entirely encoded into warp factors (as opposed to deforming the initial CY to a non-K\"{a}hler manifold).\footnote{For a comprehensive review of flux compactfications, see \cite{Grana}.} In this context there is a lot of interest in the KKLT proposal for dS vacua in type IIB with stabilized moduli \cite{KKLT}. However, the uplifting of the original AdS vacuum to dS has turned out to be challenging. Recently it was realized that a natural way to achieve this is provided by using metastable dynamical supersymmetry breaking (MDSB). Namely, by coupling the KKLT sector to an ISS sector\footnote{See \cite{ISS} for more details on the Seiberg dual of SQCD, which is now usually called the ISS model.}, one obtains a dS vacuum that is a result of spontaneous supersymmetry breaking and so is under control in the effective supergravity description \cite{DPP,AHKO}.

It was argued in \cite{KL} that the essential features of F-term uplifting in the KKLT model due to MDSB can actually be captured by taking the uplifting sector to be simply the O'Raifeartaigh model. The resulting so called O'KKLT model is significantly more tractable than the one with ISS uplifting. It was also pointed out there that considering two exponents in the non-perturbative superpotential, instead of one as in the original KKLT proposal, is beneficial for alleviating the tension between low scale supersymmetry breaking and the standard high scale cosmological inflation. Motivated by this, we studied in \cite{AC} the effective potential of the O'KKLT model at finite temperature.

Thermal corrections to a model exhibiting MDSB (more precisely, the ISS model) were first studied in \cite{ACJK,FKKMT}. These works showed that the metastable minimum is thermodynamically preferable compared to the global supersymmetric vacuum. Thus the finite temperature considerations provide a natural mechanism for the system to end up at zero temperature in a local minimum instead of in the global one. The same conclusion holds also upon coupling the ISS model to supergravity \cite{ART}. However, understanding the finite temperature behaviour of the full KKLT-ISS model is rather complicated technically.\footnote{This problem was studied in the recent work \cite{CP}.} Investigating the O'KKLT model instead, we are able to compute the one-loop temperature corrections to the effective potential for the volume modulus $\rho$ and study the resulting phase structure of the system.

\section{O'KKLT model}

The O'KKLT model of \cite{KL} is determined by the following K\"{a}hler potential and superpotential:
\be \label{OKKLT}
K = - 3 \ln (\rho + \bar{\rho}) + S \bar{S} - \frac{(S \bar{S})^2}{\Lambda^2} \, , \qquad W = W_0 + f(\rho) - \mu^2 S \, ,
\ee
where the function $f$ is either
\be \label{ftwo}
f(\rho) = A e^{-a \rho} \qquad {\rm or} \qquad f(\rho) = A e^{-a \rho} + B e^{-b \rho} \, .
\ee
In other words, this is the combination of the KKLT model (with volume modulus $\rho$) and the O'Raifeartaigh model. In fact, the latter is considered in the regime in which the two heavy scalar fields are integrated out and one is left with a single field, $S$. The last term in $K$ comes from the leading one-loop correction in an expansion in $\frac{\lambda^2 S \bar{S}}{m^2} <\!\!< 1$, where $m$ and $\lambda$ are the remaining couplings in the full O'Raifeartaigh superpotential: $m \phi_1 \phi_2 + \lambda S \phi_1^2 -\mu^2 S$. The parameter $\Lambda$ in $K$ denotes a particular combination of couplings, namely $\Lambda^2 = \frac{16 \pi^2 m^2}{c \lambda^4}$ with $c$ being a numerical constant of order 1. Also, we assume $m, \mu, \Lambda <\!\!< 1$ and consider the field space region where $S \bar{S} <\!\!< m^2/\lambda^2 <\!\!< 1$ (we work in units $M_P=1$). For more details, see \cite{KL,AC}. We will study the effective model given by (\ref{OKKLT})-(\ref{ftwo}), regardless of any underlying microphysics.

At zero temperature the KKLT model alone has one or two AdS vacua, depending on whether there are one or two exponents in $W$, that are situated at finite values of $\rho$. The O'Raifeartaigh model uplifts one of this minima to dS.\footnote{Clearly, the Dine-Seiberg minimum at infinity is also present.} Our goal is to study the phase structure of the theory (\ref{OKKLT}) at finite temperature in order to answer the question whether the fields roll towards the dS vacuum upon cooling down. 

\section{Effective potential at finite temperature}
\setcounter{equation}{0}

We will compute the one-loop finite temperature effective potential for the O'KKLT model. Let us begin by recalling the general expression for a field theory of a set of fields $\{ \chi^A \}$. It was derived first for a renormalizable theory in \cite{DJ}, using the zero-temperature functional integral method of \cite{Jackiw}, and later generalized to supergravty in \cite{BG1}. Namely, the effective potential is a function of the background values of the fields $\{\hat{\chi}^A\}$, which to one-loop order is given by: 
\be
V_{eff} (\hat{\chi}) = V_{tree} (\hat{\chi}) + V_0^{(1-loop)} (\hat{\chi})
+ V_T^{(1-loop)} (\hat{\chi}) \, .
\ee
Here $V_{tree}$ is the standard classical $N=1$ supergravity expression:
\be \label{pot}
V_{tree} = e^K (K^{A\bar{B}} D_A W D_{\bar{B}} \overline{W} - 3 |W|^2) \, ,
\ee
$V_0^{(1-loop)}$ is the zero temperature one-loop contribution, encoded in the Coleman-Weinberg
formula, and finally the temperature-dependent correction is:
\be \label{Texpansion}
V_T^{(1-loop)}(\hat{\chi}) = -\frac{\pi^2 T^4}{90}
\left( g_B+ \frac{7}{8} g_F \right)+ \frac{T^2}{24} \left[{\rm Tr}
M_s^2(\hat{\chi}) + {\rm Tr} M_f^2(\hat{\chi}) \right] + {\cal O}(T) \, .
\ee
In the above formula, ${\rm Tr} M_s^2$ and ${\rm Tr} M_f^2$ denote the traces over the mass matrices of the scalar and fermion fields respectively in the classical background $\{\hat{\chi}^A\}$\footnote{In (\ref{Texpansion}), ${\rm Tr} M_f^2$ is computed summing over Weyl fermions.}, which are given by \cite{BG2}: 
\be \label{TrMs}
{\rm Tr} M^2_s = 2 \,\la K^{C\bar{D}} \,\frac{\partial^2 V_{tree}}{\partial \chi^C
\partial \bar{\chi}^{\bar{D}}} \,\ra 
\ee
with $V_{tree}$ being the same as in (\ref{pot}) and
\be \label{TrMf}
{\rm Tr} M_f^2 = \la e^G \left[ K^{A \bar{B}} K^{C \bar{D}} (\nabla_A G_C
+ G_A G_C) (\nabla_{\bar{B}} G_{\bar{D}} + G_{\bar{B}}
G_{\bar{D}}) - 2 \right] \ra \, ,
\ee
where $G = K + \ln |W|^2$. The constants $g_B$ and $g_F$ in (\ref{Texpansion}) are the total numbers of bosonic and fermionic degrees of freedom. Since we will only need the derivatives of $V_{eff}$ w.r.t. some $\hat{\chi}^A$, we will drop from now on the $T^4$ piece of the effective potential. Also, for convenience, we will denote the remaining expression in (\ref{Texpansion}) just by $V_T$. Let us also note that the high temperature expansion (\ref{Texpansion}) is valid only in the regime, in which all masses are much smaller than the energy scale set by the temperature. 

As in \cite{KL}, we will study the effective potential in the following classical background:
\be \label{BGR}
\la \rho \ra = \langle \bar{\rho} \rangle = \sigma \, , \qquad \langle S 
\rangle = \langle \bar{S} \rangle = s \, .
\ee
Since $S$ is small in the field space region that we consider, we can expand the potential in powers of $s$. For detailed computations and expressions up to order $s^3$, see \cite{AC}.

\section{Phase structure at finite T} \label{PhaseStr}
\setcounter{equation}{0}

It is usually expected that at high temperature the effective potential is dominated by its temperature-dependent part $V_T$. So the system under consideration is naturally in a minimum of the latter. As the temperature decreases, a point may be reached at which the system undergoes a second order phase transition and starts rolling towards a different minimum. In order to investigate the details of this picture for the O'KKLT model, we first have to address the question whether there are finite vev minima of the relevant potential $V_T$. In \cite{AC} we show that for $f(\rho) = A e^{-a \rho}$ in (\ref{OKKLT}) the behaviour of $V_T$ is in fact runaway, implying decompactification of the internal space. On the other hand, for $f(\rho) = A e^{-a \rho} + B e^{-b \rho}$ the temperature-dependent part $V_T$ {\it can} have minima at finite vevs. So in the rest of this section we concentrate on the case of two exponents in the superpotential. 

In this case, the potential $V_T$ generically has two minima with finite field vevs. Assuming that the starting point at high temperature is one of those minima (we take the deeper one, which is situated at lower vevs), one can find the critical temperature $T_c$ for a second order phase transition, together with the field space position $\hat{\chi}_c$ at which the transition occurs, by solving the system
\be \label{Tcxc}
V'_{eff} (T_c, \hat{\chi}_c) = 0 \qquad {\rm and} \qquad V''_{eff} (T_c, \hat{\chi}_c) = 0 \, ,
\ee
where we have symbolically denoted by $'$ and $''$ first and second derivatives with respect to $\hat{\chi}$. We want to know whether as a result of this transition the system starts rolling towards the zero-temperature dS vacuum or in the opposite direction.\footnote{It turns out that the finite $T$ minimum, which is our starting point, is always between the dS and the susy  $T\!=0$ vacua \cite{AC}.} Let us first note that for a system with $\hat{\chi} \rightarrow - \hat{\chi}$ symmetry the above equations reduce to the more familiar condition $V_T'' = - m^2$, where $m$ is the zero-temperature mass of the field whose nonzero vev characterizes the new vacuum. In our case, however, there is no such simplification and we need to solve the full system (\ref{Tcxc}). Unfortunately though, these are transcendental equations that cannot be solved analytically (for more details, see \cite{AC}). So we turn to numerical considerations. 

Before proceeding further, we have to make a conceptual remark. Clearly, the numerical results will depend on the values of the various parameters in (\ref{OKKLT})-(\ref{ftwo}).\footnote{One can take $A=1$ without any loss of generality and then the essential parameters are $b/a$, $B$, $W_0$, $\mu$ and $\Lambda$ \cite{AC}.} In particular, for some choices of parameters there may be a single minimum of $V_T$ at finite vevs or even no minimum at all (of course, other than the Minkowski one at infinity). This is the same kind of situation as for the zero-temperature potential $V_0$ studied in \cite{KL}. Namely, there are many parameter values such that $V_0$ does not have a dS (or even any) minimum. What is important, however, is that there exist many parameter choices for which $V_0$ does have a dS vacuum \cite{KL}. These are exactly the parameter values that are relevant for moduli stabilized dS vacua. In the same vein, here we are interested in the sets of parameters for which $V_T$ has at least one finite vevs minimum, {\it when} $V_0$ has a dS vacuum, as these are the choices for which one can have a dS compactification that is not destabilized by thermal effects.

We have found several sets of parameters, for which the system is in the desired regime; see Table \ref{table}, where $x=a\sigma$.
\begin{table}[h]
\begin{center}
\vspace{0.4cm}
\begin{tabular}{|c|c|c|c|c|c|c|} 
\hline
 $B$ & $W_0$ & $\mu$  & $\Lambda$ & $x_{dS}^{(0)}$ & $x_{AdS}^{(0)}$ & $x_{min}^{(T)}$ \\ 
\hline 
-1.040 & $-7.6 \times 10^{-5}$ & $8 \times 10^{-4}$ & $10^{-2}$ & 4.88 & 7.84 & 5.62 \\
\hline
-1.036 & $-1.1 \times 10^{-4}$ & $2 \times 10^{-3}$ & $10^{-2}$ & 4.50 & 7.40 & 5.25 \\
\hline  
-1.032 & $-1.64 \times 10^{-4}$ & $ 10^{-3}$ & $10^{-2}$ & 4.11 & 6.92 & 4.83 \\ 
\hline
-1.028 & $-2.4 \times 10^{-4}$ & $0.66 \times 10^{-3}$ & $10^{-3}$ & 3.73 & 6.44 & 4.44 \\ 
\hline
-1.024 & $-3.533 \times 10^{-4}$ & $0.66 \times 10^{-3}$ & $10^{-3}$ & 3.34 & 6.00 & 4.04 \\
\hline
-1.020 & $-5.21 \times 10^{-4}$ & $0.95 \times 10^{-3}$ & $10^{-3}$ & 2.96 & 5.52 & 3.64 \\
\hline
-1.016 & $-7.67 \times 10^{-4}$ & $1.4 \times 10^{-3}$ & $10^{-2}$ & 2.55 & 5.02 & 3.20 \\
\hline 
\end{tabular}
\end{center}
\vspace{-0.2cm}
\caption{ {\small Each row represents a set of parameters for which both $V_0$ and $V_T$ have minima at finite field vevs {\it and} the lower-$x$ minimum of $V_0$ is dS. In each set $b/a=100/99$ as in the examples of \cite{KL}. The positions of the minima are denoted by $x_{dS}^{(0)}$ and $x_{AdS}^{(0)}$ for $V_0$ and by $x_{min}^{(T)}$ for the lower-$x$ minimum of $V_T$. Note also that $x=a\sigma$ and so, taking $a=\frac{\pi}{100}$ for instance, the various minima occur for $\sigma \sim {\cal O} (100)$.} \label{table}}
\end{table}
According to this table, the O'KKLT model seems to exhibit the behaviour of interest only at discrete points in parameter space. Strictly speaking, this is only true up to a small variation of one (or more) parameter(s) for some of the sets; for more details see \cite{AC}. Nevertheless, it is quite interesting to note that in order to move from one point in parameter space to another, such that at both points one has the same kind of physics, generically one needs to change in a discrete way at least two parameters. This suggests that anthropic arguments of the kind of \cite{SW} might be too naive. Recall that in such considerations one usually argues that a particular value of a given constant of nature (for example, the cosmological constant) is anthropically preferred since varying that constant, while keeping all other coupling constants fixed, leads to drastic physics changes. Given the example of the O'KKLT model however, it seems conceivable that these anthropic/environmental arguments might fail under more general variations, i.e. discrete changes of more than one coupling constant at the same time.\footnote{However, see \cite{Hamed} for arguments in favor of Weinberg's argument in the case when both the cosmological constant and the Higgs mass are varied, while the remaining couplings are held fixed.} This issue is well-worth investigating in more realistic models.

\begin{figure}[t]
\begin{center}
\scalebox{0.72}{\includegraphics{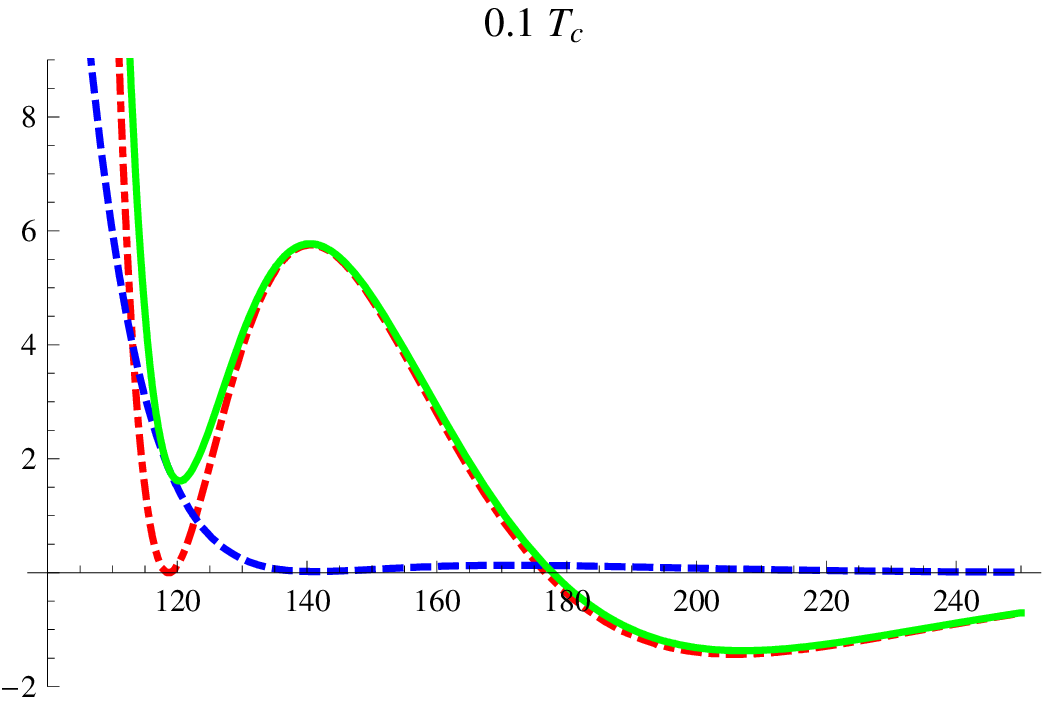}} \hspace*{0.4cm}\scalebox{0.72}{\includegraphics{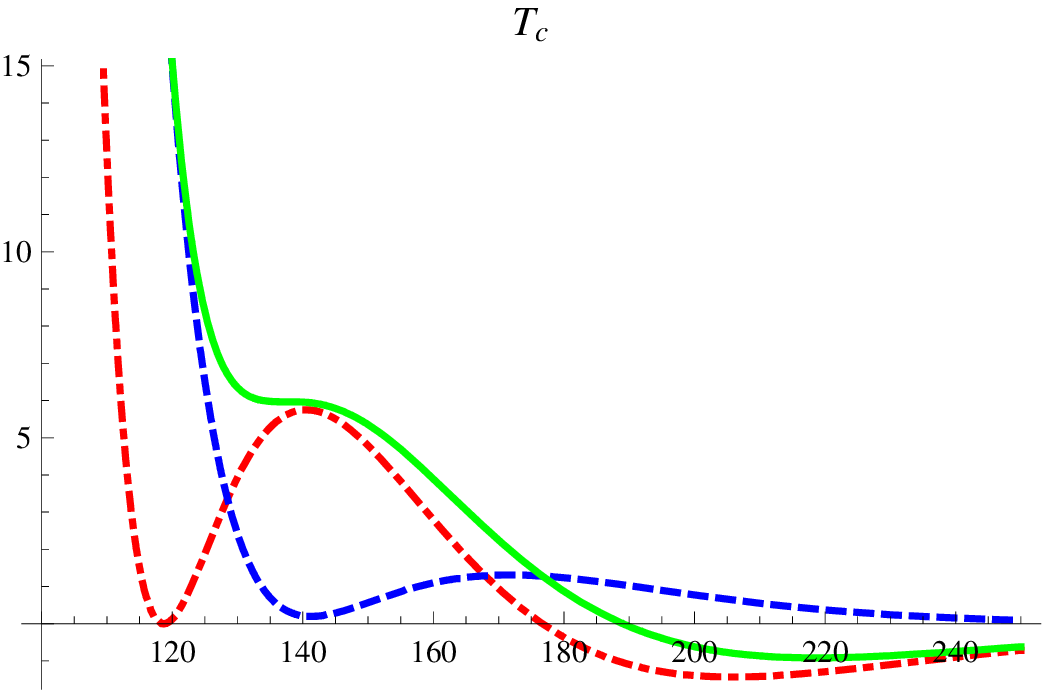}}
\end{center}
\vspace{-0.5cm}
\caption{ {\small Effective potential $V_{eff}$ (green continuous line), multiplied by $10^{15}$, as a function of $\sigma$ for the race-track type model compared to $V_0 (\sigma)$ (red dot-dashed line) and $V_T (\sigma)$ (blue dashed line) for $T=0.1 \times T_c$ (left) and for 
$T= T_c$ (right). The values of the parameters are the following: $a=\pi/100$, $B = -1.028$, $b/a=\frac{100}{99}$, $W_0=-2.4\times 10^{-4}$, $\mu = 0.6 
\times 10^{-3}$, $\Lambda=10^{-3}$; the resulting critical temperature is $T_c = 0.3$.}}
\label{OKLTc}
\end{figure}
Let us now turn to the question about the occurrence of a second order phase transition. From Table \ref{table} we see that the $V_T$ minimum of interest is situated between the dS and AdS vacua of the zero temperature potential. So it seems meaningful to ask whether the system rolls towards the meatstable or the supersymmetric $T=0$ vacua upon cooling down. However, the critical temperature for the relevant second order phase transition turns out to be always of order $0.1$; an example, representative for {\it all} rows of Table \ref{table}, is illustrated in Figure \ref{OKLTc}.\footnote{For convenience, the graph of $V_{eff}$ (the green continuous line) does not include the term $\sim T^4$ in (\ref{Texpansion}); as the latter is field-independent, it only leads to an irrelevant overall shift of the $V_{eff}$ plot.} Now, recall that we are working in units of $M_P =1$. Hence the critical temperature that we obtain is $T_c \sim {\cal O}(0.1 \,M_P)$. This is rather unexpected since the potential barrier between the relevant minima is of order $10^{-15} M_P$, as can be seen on the figure, and so one would have expected intuitively that $T_c <\!\!< M_P$. In fact, such a high $T_c$ implies that we cannot reliably conclude that there is a phase transition, since the supergravity approximation is near the threshold of its validity. Furthermore, such $T_c$ is clearly outside of the validity of the single field approximation to the O'Raifeartaigh model.\footnote{Recall that this low-energy effective description of the O'Raifeartaigh model, obtained by integrating out the two heavy fields, is a motivation to consider the model determined by (\ref{OKKLT})-(\ref{ftwo}), although the latter can certainly be studied on its own without viewing it as an approximation to the O'Raifeartaigh model (since neither of the two is supposed to provide a fundamental description anyway).} 
Nevertheless, we can turn this around and conclude that for the whole range of validity of the supergravity approximation, i.e. for $T <\!\!< M_P$, (and hence even more so for the whole range of validity of the single field low-energy approximation to the full O'Raifeartaigh model) the extrema of the effective potential of the system are determined by the zero-temperature part and not by $V_T$.\footnote{Note that, because of the constant term $\sim T^4$ that we are omitting, this does not mean that the magnitude of $V_{eff}$ itself is determined by $V_0$.} Therefore, at the level of our approximations the $T=0$ de Sitter minimum is not destabilized by thermal effects. 

If we take the O'KKLT model as a simple toy model of the early Universe, the behaviour that we have uncovered above implies that reheating does not destabilize the metastable minimum of the zero-temperature potential $V_0$ for the volume modulus $\sigma$. Indeed, at the end of the inflationary stage the Universe is generically expected to be very cold. Hence it will be in a local minimum of $V_0$.\footnote{Note that the modulus $\sigma$, defined in (\ref{BGR}), {\it should not} be confused with the inflaton field.} The exit from inflation comes with the decay of the inflaton into various particles and the resulting reheating of the Universe to some temperature $T_R$. It is usually expected that $T_R<\!\!< M_P$ and so our considerations can be applied to this system, leading to the conclusion that we stated in the beginning of this paragraph. Namely, we can conclude that reheating does not destabilize the dS vacuum of the volume modulus or, in other words, does not lead to decompactification of the internal space.

This new point of view raises the question whether the zero-temperature dS vacuum could remain a local minimum of the total effective potential even for parameter values for which $V_T$ has a runaway behaviour (of course, within the range of validity of our approximations). It is easy to verify that this is indeed the case for sets of parameters that are small variations of those in Table \ref{table}, such that $V_T$ does not have any finite-vev minima while the dS minimum of $V_0$ is preserved. A more detailed study of this issue for a broad range of parameter values merits a separate investigation.

\section{One exponential revisited} \label{OneExp}
\setcounter{equation}{0}

In light of the results of the previous section, it is worth to re-examine the case of one exponent in the superpotential in order to see whether the zero-temperature dS vacuum can survive at high $T$, despite the runaway behaviour of $V_T$ in this case. Let us, for simplicity, view here the model (\ref{OKKLT}) with $f(\rho) = A e^{-a \rho}$ as an effective description on its own.\footnote{I.e., we will only worry about one cut-off scale - the Planck scale.} One can show analytically \cite{AC} that the dS minimum of $V_0$ is not washed out by the thermal corrections only when certain order-of-magnitude inequalities are satisfied. In particular, we have found that for the largest allowed values of the parameters $W_0$ and $\mu$ (i.e., the values that require the smallest amount of fine-tuning and are thus the most preferable ones) the $T=0$ dS minimum persists in the total $V_{eff}$ as long as $T<\Lambda$. Clearly then, by taking $\Lambda \sim {\cal O}(10^{-2})$ one can ensure that the dS vacuum is not destabilized by the temperature contribution for the whole range of validity of the supergravity approximation. Obviously, for $\Lambda \sim {\cal O}(10^{-3})$ or smaller this is not the case anymore as temperatures greater than $\Lambda$ are well within the range $T<\!\!<M_P$ and hence one can reliably conclude that at high $T$ the internal space decompactifies. 

It is interesting to note that the recent work \cite{BCCN} reached the same conclusions, that we did in Sections \ref{PhaseStr} and \ref{OneExp} here, from a different perspective. Namely, they studied the dynamics of a similar system at finite temperature and showed that the thermal corrections have quite a limited effect on its evolution and that, under certain conditions, one can have temperatures as high as $10^{-2}M_P$ without losing the minimum of the zero-temperature potential, despite the lack of finite vev minima of $V_T$.

\section*{Acknowledgements}
L.A. would like to thank the organizers of the 3rd RTN Workshop "Constituents, Fundamental Forces and Symmetries of the Universe" held in Valencia, October 2007, for the opportunity to present this work. The research of L.A. is supported in part by the EU RTN network MRTN-CT-2004-005104 and INTAS contract 03-51-6346. The work of V.C. is supported by a Queen Mary studentship.

\end{document}